\def\bold#1{\setbox0=\hbox{$#1$}%
      \kern-.02em\copy0\kern-\wd0
      \kern.04em\copy0\kern-\wd0
      \kern-.02em\raise.0433em\box0 }

\def\bdsmall#1{\setbox0=\hbox{$#1$}%
      \kern-.015em\copy0\kern-\wd0
      \kern.03em\copy0\kern-\wd0
      \kern-.015em\raise.0233em\box0 }

\documentstyle[12pt]{article}
\begin{document}
\rm
\centerline{\bf{A TEST OF THE EIKONAL APPROXIMATION}}
\centerline{\bf{IN HIGH--ENERGY (e,e$'$p) SCATTERING}}
\vskip1.2truecm
\centerline{\small{A.~Bianconi and M.~Radici}}

\centerline{\small \it{ Dipartimento di Fisica Nucleare e Teorica,
Universit\`a di Pavia, and}}
\centerline{\small \it{ Istituto Nazionale di Fisica Nucleare,
Sezione di Pavia,}}
\centerline{\small \it{ v. Bassi 6, 27100 Pavia, Italy}}

\vskip2.truecm

\begin{abstract}

The Glauber method is extensively used to describe the motion of a hadronic
projectile in interaction with the surrounding nuclear medium. One of the
main approximations consists in the linearization of the wave equation for
the interacting particle. We have studied the consequences of such an
assumption
in the case of the $^{12}\mathrm{C(e,e}'\mathrm{p)}^{11}\hbox{\rm B}^*$
reaction at high proton momenta by comparing the results with the predictions
obtained when all the ingredients of the calculation are unchanged but the
second-order differential equation for the scattered wave, which is solved
exactly for each partial wave up to a maximum of 120 spherical harmonics.
We find that the Glauber cross section is always larger by a factor $10 \div
20 \%$, even at vanishing missing momenta. We give a quantum-mechanical
explanation of this discrepancy. Nevertheless, a good correlation is found
between the two predictions as functions of the missing momentum, especially
in parallel kinematics.

\end{abstract}

\vskip 1.5cm

The problem of the validity of the Glauber approximation~\cite{glauber} in
(e,e$'$p)
scattering from finite nuclei is receiving more and more interest in relation
with the experiments
planned at CEBAF, where the proton momentum can be larger than $1 \div 2$
GeV/c. At large missing momenta of the recoil the details of the short-range
nucleon-nucleon interaction are expected to show up~\cite{theory}, but the
few experimental data available~\cite{exp}
still prevent from putting stringent constraints on the
theoretical models. On the other side, at moderate missing momenta an
accuracy within $10\%$ is required to unambigously identify exotic effects
like Colour Transparency, if any~\cite{CT}.

The recent NE18 experiment~\cite{NE18exp} has shown that calculations based
on the standard
Glauber approximation overestimate the cross section even at vanishing
missing momenta, unless some phenomenological corrections are introduced.
Various suggestions have been made both on the way of analyzing the NE18
data~\cite{NE18CT,pperp} and on how to improve the Glauber
method~\cite{Pandh}.
Starting from a different point of
view, we show that the latter is affected by an intrinsic systematical error,
which can be relevant at the kinematics explored at CEBAF but seems anyway
to be predictable and perhaps can be corrected for.

Traditionally, the Glauber method has been extensively used in the analysis
of the proton-nucleus scattering data~\cite{bw68,abv78}, where it gives good,
or at least some of
the best available, fits to the scattering distributions. However, the
generalization to the (e,e$'$p) reaction is not straightforward. In fact,
even if the hadronic content of the final state is similar, the inelastic
nature of the electromagnetic proton emission makes the kinematical situation
rather
different.

The basic ingredient to describe the knockout process for an exclusive
scattering in the framework of DWIA, is the scattering
amplitude~\cite{fm84,bgp93}
\begin{equation}
J^{\mu}_{\alpha} ({\bold q}) = \int \hbox{\rm d} {\bold r} \hbox{\rm d}
\sigma \hbox{\rm e}^{\mathrm{i}
{\bdsmall {\scriptstyle q}} \cdot {\bdsmall {\scriptstyle r}}}
\chi^{\left( -\right)\, *} ({\bold r}, \sigma) {\hat J}^{\mu} ({\bold q},
{\bold r}, \sigma) \Psi_{\alpha} ({\bold r}, \sigma) \quad ,
\label{eq:scattampl}
\end{equation}
where ${\bold q}$ is the momentum transfer, $\chi^{\left( - \right)}$ is the
distorted wave function of the knockout nucleon and $\Psi_{\alpha}$ is the
single-particle bound state wave function depending on the quantum numbers
$\alpha$ of the hole. For sake of simplicity, in this work we have focussed
on
the longitudinal component of the current operator, retaining just the
leading order $O(1)$ in the nonrelativistic expansion and neglecting the
nucleon form factor, i.e. substituting ${\hat J}^{\mu}$ with the identity
operator. The cross section becomes, therefore, proportional to
\begin{equation}
\Big \vert \int \hbox{\rm d} {\bold r} \hbox{\rm d} \sigma \hbox{\rm e}^{
\mathrm{i}
{\bdsmall { \scriptstyle q}} \cdot {\bdsmall {\scriptstyle r}} } \chi^{\left(
- \right) \, *} ({\bold r}, \sigma) \Psi_{\alpha} ({\bold r}, \sigma) \Big
\vert^2 \equiv S^{\mathrm{D}}_{\alpha} ({\bold q}) \quad ,
\label{eq:specdist}
\end{equation}
which can be identified as the "distorted" spectral density
$S^{\mathrm{D}}_{\alpha}$~\cite{bgp82} at the missing energy corresponding
to the knockout hole $\alpha$.

The distorted wave function $\chi^{\left( - \right)}$ is solution of the
Schr\"odinger equation
\begin{equation}
\left( - {\hbar^2 \over {2 m}} \nabla^2 + V \right) \chi = E_{
\mathrm{cm}} \chi \quad ,
\label{eq:schroeq}
\end{equation}
where $m$ is the reduced mass of the nucleon in interaction with the residual
nucleus, $E_{\mathrm{cm}}$ is its kinetic energy in the cm
system and $V$ in principle includes a nonlocal energy-dependent optical
potential effectively describing the residual
interaction. The incoming-wave boundary conditions to be imposed on the
solution of eq. (\ref{eq:schroeq}) correspond to the well known feature of
the quantum-mechanical problem for a
wave crossing a potential barrier, where the asymptotic stationary
conditions require the incoming unitary flux to be splitted in a reflected
one and in an outgoing one.

The proper way of solving eq. (\ref{eq:schroeq}) is to expand
$\chi^{\left( - \right)}$ in partial waves, solve the second-order
differential equation wave by wave, sum the solutions up to a certain maximum
angular momentum $L_{\mathrm{max}}$ which satisfies an {\it ad hoc}
convergency
criterion for eq. (\ref{eq:scattampl}). From now on, this procedure will be
referred to as method I. So far, it has been applied
to proton momenta below $0.3$ GeV/c~\cite{bgp93}, where convergency is
reached for $L_{\mathrm{max}} < 50$. In this work we have extended it
to larger proton momenta, i.e. $p \sim 1 \div 2$ GeV/c. A maximum
$L_{\mathrm{max}} = 120$ has been used and we have checked that the results
become already stable with $L_{\mathrm{max}} = 100$.

At high energies the Glauber method~\cite{glauber} offers an alternative way
(from now on
method II) of solving eq. (\ref{eq:schroeq}) by linearizing it along the
propagation axis $\hat z$:
\begin{eqnarray}
{\bold r} &\equiv& z {\displaystyle {{\bold p} \over p}} + {\bold b}
\label{eq:zb} \\
\nabla^2 &\simeq& {\displaystyle {\partial^2 \over {\partial z^2}}}
\label{eq:nabla} \\
\left( {\partial^2 \over {\partial z^2}} + p^2 \right) &=& {\displaystyle
\left( {\partial \over {\partial z}} + \hbox{\rm i} p \right) \cdot \left(
{\partial \over {\partial z}} - \hbox{\rm i} p \right)} \nonumber \\
&\simeq& {\displaystyle 2 \hbox{\rm i} p \cdot \left( {\partial \over
{\partial z}} - \hbox{\rm i} p \right)} \quad ,
\label{eq:glau}
\end{eqnarray}
where ${\bold b}$ is the impact parameter describing the degrees of freedom
transverse to the motion of the struck particle with momentum ${\bold p}$.
With this approximation eq. (\ref{eq:schroeq}) becomes
\begin{equation}
\left( {\partial \over{\partial z}} - \hbox{\rm i} p \right) \chi = {1 \over
{2 \mathrm{i} p}} V \chi \quad .
\label{eq:schroglau}
\end{equation}
The standard boundary condition applied requires that asymptotically $\chi
\rightarrow 1$, which corresponds to an incoming unitary flux of plane
waves. In contrast to method I no reflected flux is taken into account that
is created by the potential of the nucleon-nucleon interaction. Because of
the different phases, there is a destructive interference between the
reflected and the incoming waves, which eq. (\ref{eq:schroglau}) neglects
thus leading to an overestimation of eq. (\ref{eq:specdist}). In particular,
the approximation
of eq. (\ref{eq:glau}), upon which the eq. (\ref{eq:schroglau}) is based,
can be expected to produce most of the discrepancy at very small angles
$\gamma$ between ${\bold q}$ and ${\bold p}$, typically in parallel
kinematics.
On the other side, at large
$\gamma$ and large transverse missing momenta the approximation of eq.
(\ref{eq:nabla}) is expected to play the major role.

A direct comparison with the exact solutions of eq. (\ref{eq:schroeq}),
summed up
to $L_{\mathrm{max}}$, is needed to get a quantitative answer out of these
qualitative considerations. The need for considering corrections beyond the
eikonal approximation of the Glauber method is a well known problem in
high-energy elastic proton scattering on nuclei~\cite{abv78,wall75,rin76}.
One basic outcome of these studies is that competing effects tend to cancel
each other at small deflection angles~\cite{wall75}, while can give important
corrections at larger angles~\cite{rin76}. This fact suggests that also in
(e,e$'$p) reactions it could be interesting to compare deviations from the
eikonality with other kinds of corrections to the Glauber model. This is the
aim of a more general work, whose preliminary results we present in this
communication.

Here, we have considered the $^{12}\hbox{\rm C(e,e}'\hbox{\rm p)}^{11}
\hbox{\rm B}^*$ reaction both in parallel and perpendicular kinematics for
the proton momentum $p$ up to $2$ GeV/c and the momentum transfer $q = 1.4$
GeV/c. When not explicitely mentioned, the bound state $\Psi_{\alpha}$ must
be understood as the solution of the Woods-Saxon potential
of Comfort and Karp~\cite{bwav} with the quantum numbers $\alpha$ of the $s$
wave.
As an exploratory calculation, we have neglected the contribution to $V$ in
eq. (\ref{eq:schroeq}) coming from the Coulomb potential to avoid numerical
problems related to the high angular momenta required. Therefore, in proper
terms the results presented here refer to the (e,e$'$n) reaction. $V$ is an
optical potential of the form
\begin{eqnarray}
V(r) &=& \left( U + \hbox{\rm i} W \right) \, {\displaystyle {1 \over {1 +
\hbox{\rm e}^{{{r - R} \over a}}}}} \nonumber \\
&\equiv& \left( U + \hbox{\rm i} W \right) \, \rho (r) \quad ,
\label{eq:opt}
\end{eqnarray}
with $R = 1.2 \times A^{1/3}$ fm and $a = 0.5$ fm. The nuclear density $\rho
(r)$ defined in eq. (\ref{eq:opt}) is normalized such that $\rho (0) = 1$.

No phenomenological optical potential is available at the energies here
considered. Therefore, the parameters $U, W$ can only
be guessed. Two guidelines can help in this case. On one side, according to
the
Glauber model the imaginary part should scale as $W \sim p / 10$ MeV,
while $U / W$ should equal the ratio between the real and the imaginary parts
of the average proton-nucleon forward-scattering amplitude, which is expected
to be $\leq 0.5$ in the considered kinematics~\cite{lll93}. On the other
side, assuming that
the absorption due to the final state interactions, observed in the NE18
experiment in the context of a semi-inclusive (e,e$'$p), is reasonable also
for a completely exclusive knockout, then $U, W$ should be better taken about
half of the previous values. We have considered several choices, including
the unphysical but interesting cases of a completely real ($W = 0$) or
imaginary ($U = 0$) potential.

By comparing the predictions of methods I and II for eq. (\ref{eq:specdist})
at ${\bold p} = {\bold q}$ we find that
in the range $0.6 \leq p \leq 2$ GeV/c the results are $p$-independent
if $U, W$ are linear functions of $p$. This is a well known feature of the
Glauber method at high energy proton-nucleus scattering. Hence, the previous
findings suggest that, at least for small missing momenta, starting from $p
\sim 0.6$ GeV/c on we are already in a "Glauber regime". Therefore, in the
following whenever the kinematics is parallel, we employ an optical potential
of the form $(U + \hbox{\rm i} W) \rho (r) p / p_{\mathrm{o}}$, where
$p_{\mathrm{o}} = 1.4$ GeV/c.

Figs. 1 and 2 show the distorted spectral density of eq. (\ref{eq:specdist})
in parallel kinematics for missing momenta in the range $0 \leq
p_{\mathrm{m}} = p - q \leq 600$ MeV/c. The dashed line represents the result
with no final state
interaction (PWIA), which is of course identical in both methods. The solid
and the dotted lines are the results of method I and II,
respectively. The two curves are rather well correlated in all the range of
$p_{\mathrm{m}}$ here explored. But a systematic discrepancy is
evident, even at $p_{\mathrm{m}} \sim 0$, which is roughly
proportional only to the imaginary part
of the optical potential, as it can be realized by inspecting curves in fig.
1 labeled by (a) ($U = 100$ MeV, $W = 0$) and (b) ($U = 0$, $W = 100$ MeV),
or in fig. 2 by (a) ($U = 20$, $W = 50$ MeV) and (b) ($U = 50$, $W = 150$
MeV).
This fact suggests that the origin of the discrepancy should be ascribed to
the linearization of eq. (\ref{eq:schroeq}) implied by the approximation of
eq. (\ref{eq:glau}).

It is interesting also to explore the role of the Final State Interactions
(FSI), which seem to have a nontrivial structure especially at high missing
momenta in perpendicular kinematics. To accomplish this, we rewrite eq.
(\ref{eq:specdist}) as
\begin{eqnarray}
S^{\mathrm{D}}_{\alpha} ({\bold q}) &\sim& \vert \hbox{\tt PWIA} +
\hbox{\tt FSI} \vert^2 \nonumber \\
 &=& \vert \hbox{\tt PWIA} \vert^2 + \vert \hbox{\tt FSI} \vert^2 + 2
\hbox{\rm Re} ( \hbox{\tt PWIA} \cdot \hbox{\tt FSI}^* )
\label{eq:fsi}
\end{eqnarray}
and we compute the single contributions separately in the framework of method
I. In fig. 3 the $\vert \hbox{\tt PWIA} \vert^2$, $\vert \hbox{\tt FSI}
\vert^2$ and the total contribution $S^{\mathrm{D}}_{\alpha}$ are shown as
functions of the angle $\gamma$ between ${\bold q}$ and ${\bold p}$ using a
strong unrealistic optical potential ($U = 50, \, W = 150$ MeV) to
emphasize the role of FSI. At small angles the $\hbox{\tt PWIA}$ is the
dominant
contribution. The interference is negative and is responsible for the
familiar damping observed in all electron scattering data. Near $\gamma \sim
15^{\mathrm{o}}$
the angular distribution has a dip (partially filled if the real part $U$ of
the optical potential is non vanishing), because the interference is roughly
equal and opposite to the sum of $\vert \hbox{\tt PWIA} \vert^2$ and $\vert
\hbox{\tt FSI} \vert^2$. At large angles the distribution basically coincides
with $\vert \hbox{\tt FSI} \vert^2$.

To better specify this last issue let's consider the fig. 4, where the
angular distribution $S^{\mathrm{D}}_{\alpha}$ is plotted in the same
conditions as in fig. 3
but substituting inside $\Psi_{\alpha}$ the shell-model wave function of the
Woods-Saxon type with a harmonic oscillator with the same quantum numbers.
In this case the dashed line, i.e. the $\hbox{\tt PWIA}$ contribution, falls
down
monotonously and quickly just after the Fermi momentum ($221$ MeV/c at
$\gamma \sim 10^{\mathrm{o}}$). Any diffractive pattern must therefore
be ascribed to the $\hbox{\tt FSI}$ term. This remark, less obvious in the
case of fig. 3 because of the richer structure of $\hbox{\tt PWIA}$, is a
common feature of both approaches, the method I (solid line) and the method
II (dotted line). They appear correlated similarly to the case of the
parallel kinematics.

The oscillatory diffractive trend of $S^{\mathrm{D}}_{\alpha}$ at large
angles is reminiscent of the angular distribution
for proton-nucleus scattering~\cite{bw68}. Thus, the natural interpretation
is that the
ejected proton is testing coherently the residual nucleus. This is possible
only in a completely exclusive reaction, i.e. where the residual nucleus
doesn't fragment. Any measurement of an energy-integrated distribution, i.e.
of a semi-inclusive (e,e$'$p) reaction~\cite{NE18CT,pperp,deuter,helium4},
looses this information on the structure of
the $A-1$ system and tests just the average behaviour of the single ejected
nucleon.

\vskip .5cm

We have critically analyzed some aspects of the Glauber approach to the
description of the propagation of a particle through the nuclear medium
in the case of the $^{12}\hbox{\rm C(e,e}'\hbox{\rm p)}^{11} \hbox{\rm B}^*$
reaction. The basic approximation of linearizing the wave equation for the
projectile has been shown to be responsible for a systematic discrepancy with
respect to the results obtained from its proper solution. This fact could be
relevant at the CEBAF kinematics, particularly
at small missing momenta. At very large values of the transverse missing
momenta the FSI give the dominant contribution to the cross section and show
a diffractive pattern, which can naturally be interpreted as a coherent
diffractive scattering between the ejected proton and the residual nucleus.
To verify this prediction a completely exclusive (e,e$'$p) reaction is
needed, where the residual nucleus is in a well specified state.

\vskip .5cm

We would like to thank prof. S. Boffi for many discussions and for
his continuous interest in this work, and dr. F. Cannata~\cite{cdl86} for
having originally suggested the idea upon which this work is based.


\vfil \eject

\centerline{\bf Captions}

\vskip 1cm \noindent

{\bf Fig. 1} - The distorted spectral density $S_{\alpha}^{\mathrm{D}}$ as a
function of the proton momentum $p$ for the $^{12}\mathrm{C(e,e}'\mathrm{p)}
^{11}\hbox{\rm B}^*_{\mathrm{s} 1/2}$ reaction in parallel kinematics at the
momentum transfer
$q = 7 \, \mathrm{fm}^{-1}$. The dashed line is the result of the PWIA. The
solid
and the dotted lines are obtained in the framework of method I and II,
respectively (see text). The curves labelled by (a) are produced with $U=100,
\, W=0$ MeV/c, where $U, W$ are the real and the imaginary parts of the
optical potential, respectively. The curves labelled by (b) are produced with
$U=0, \, W=100$ MeV/c.

\vskip 1cm \noindent

{\bf Fig. 2} - The same as in fig. 1, but the curves labelled by (a) are
produced with $U=20, \, W=50$ MeV/c, while the curves labelled by (b) are
produced with $U=50, \, W=150$ MeV/c.

\vskip 1cm \noindent

{\bf Fig. 3} - The $S_{\alpha}^{\mathrm{D}}$ as a function of the angle
$\gamma$
between ${\bold q}$ and ${\bold p}$ for the $^{12}\mathrm{C(e,e}'\mathrm{p)}
^{11}\hbox{\rm B}^*_{\mathrm{s} 1/2}$ reaction in perpendicular kinematics at
$q = 7 \, \mathrm{fm}
^{-1}$. The dashed line is the PWIA result, the dotted line is the pure
contribution of the FSI (when the PWIA contribution is subtracted), the solid
line is the coherent sum of the two. All the curves are obtained in the
framework of method I with $U=50, \, W=150$ MeV/c.

\vskip 1cm \noindent

{\bf Fig. 4} - The same as in fig. 3 but with the bound state described by a
harmonic oscillator. The dashed line is the PWIA result, the solid and dotted
lines are the total result for method I and II, respectively.

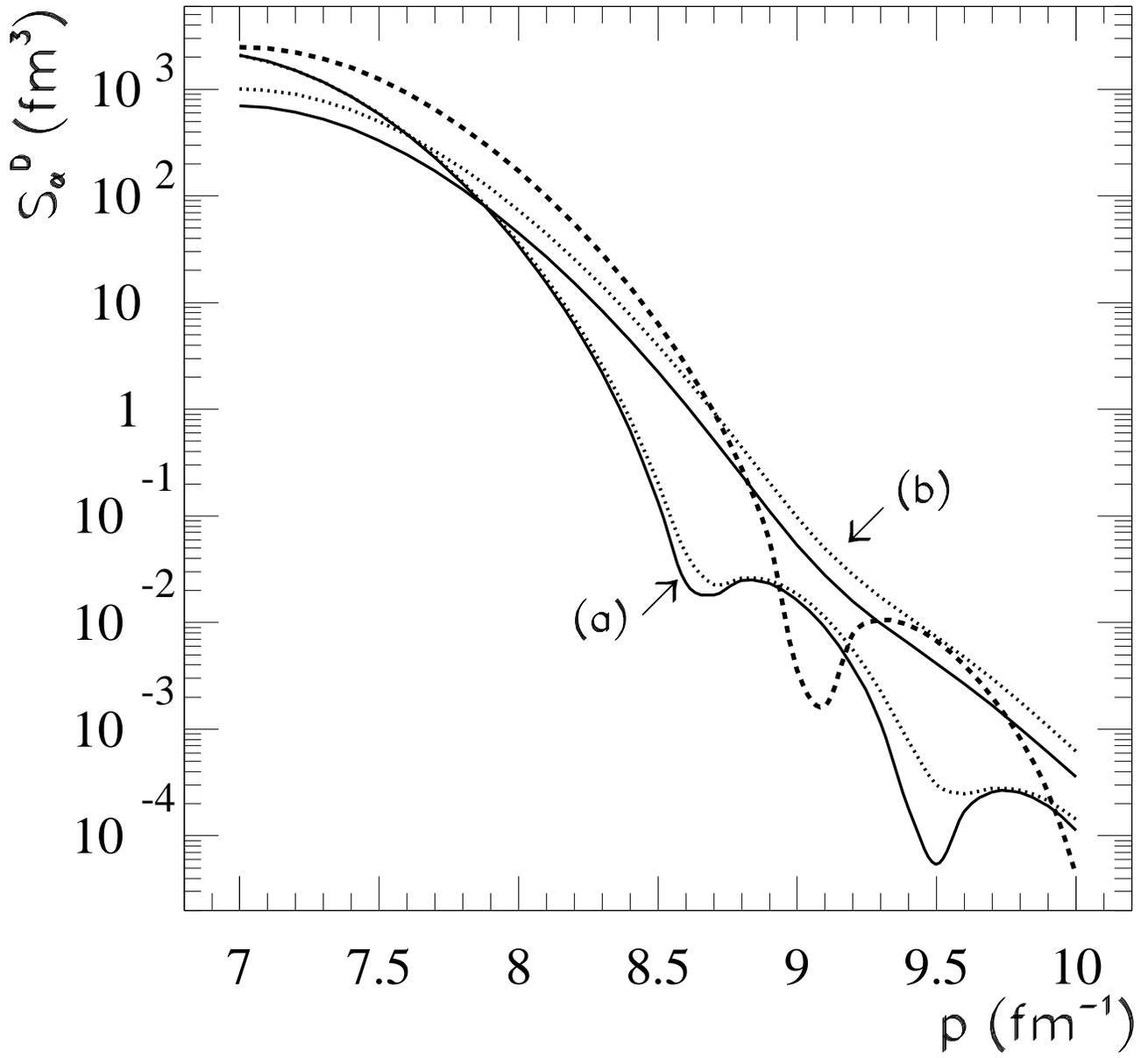
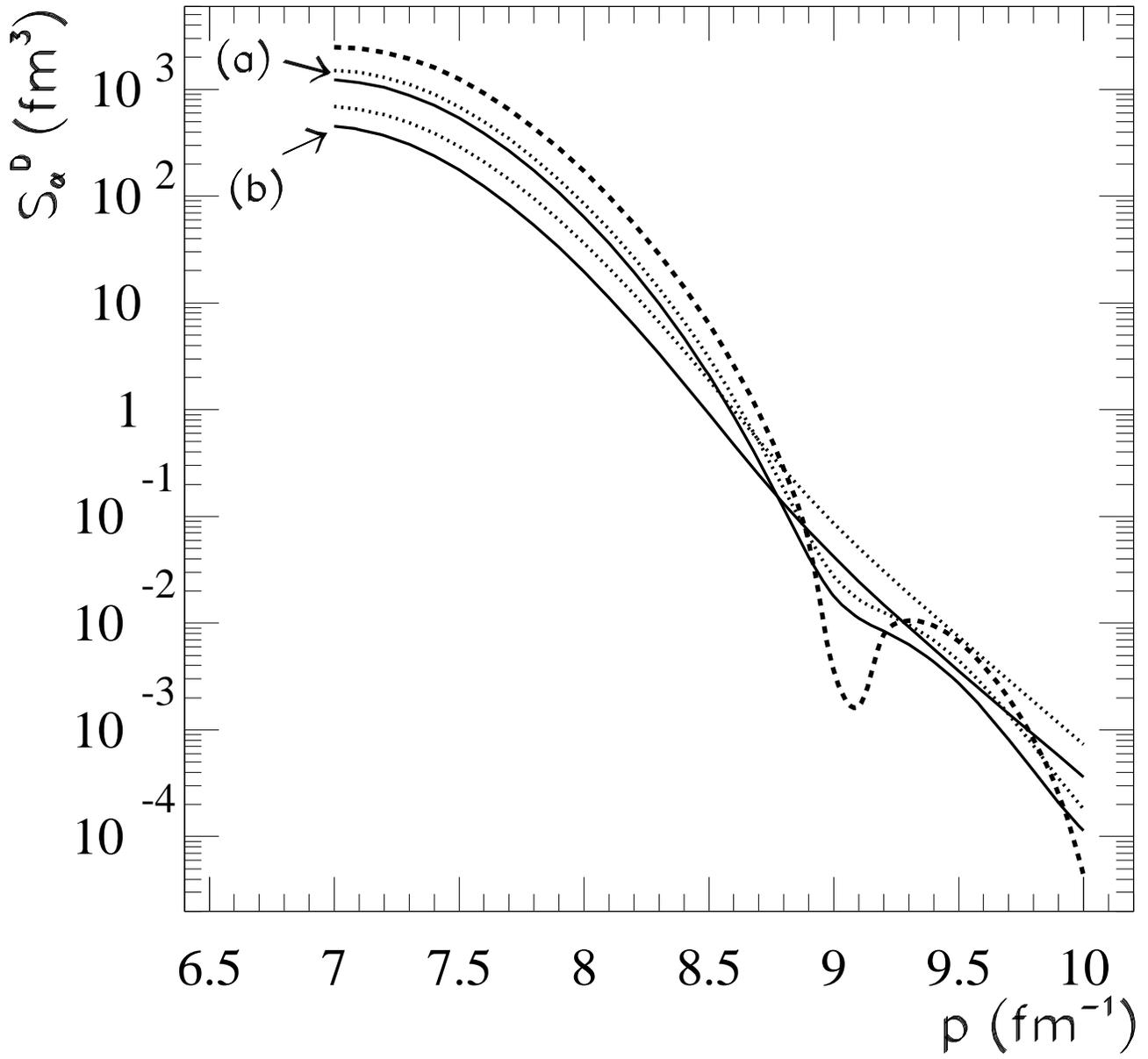
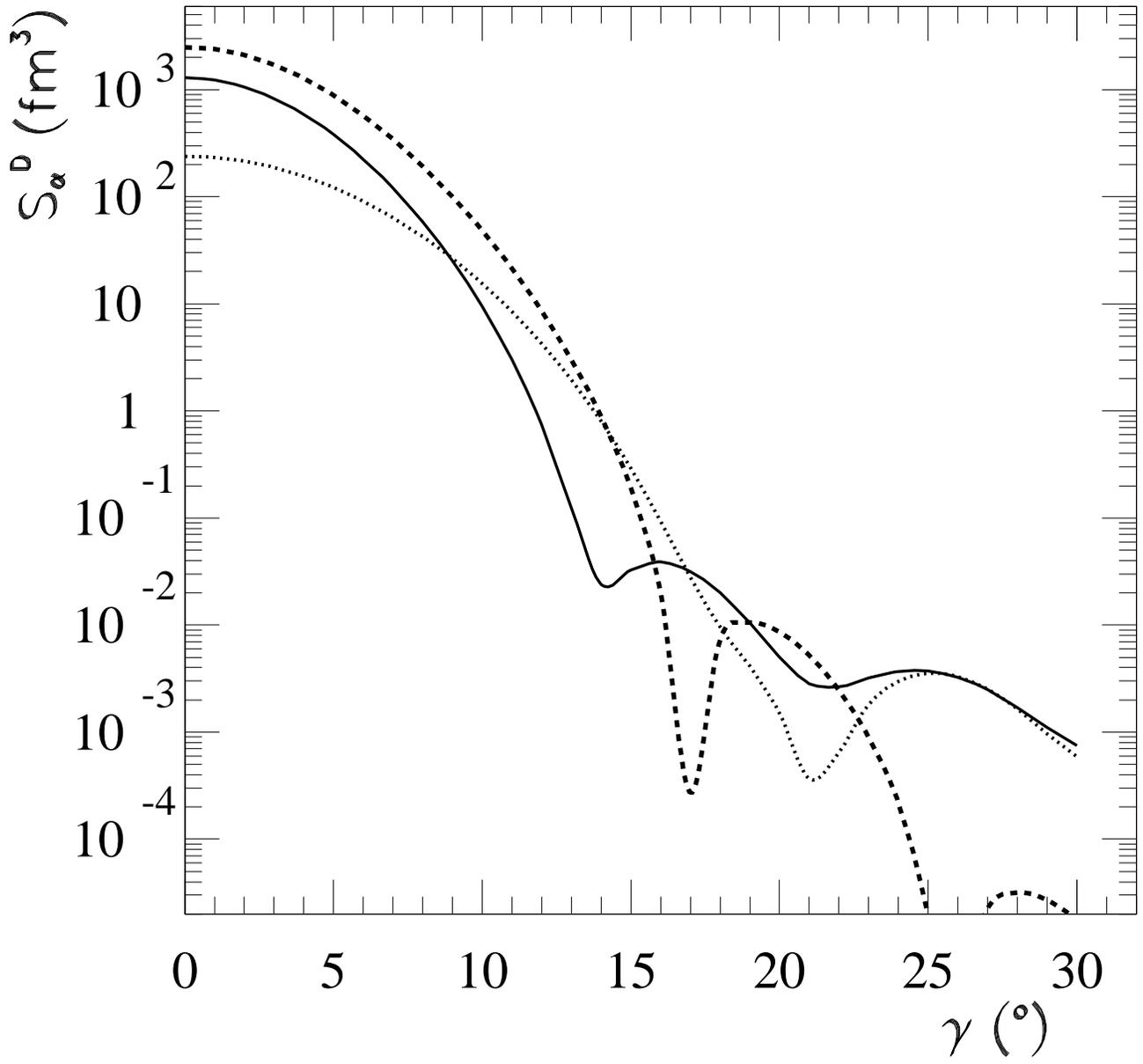
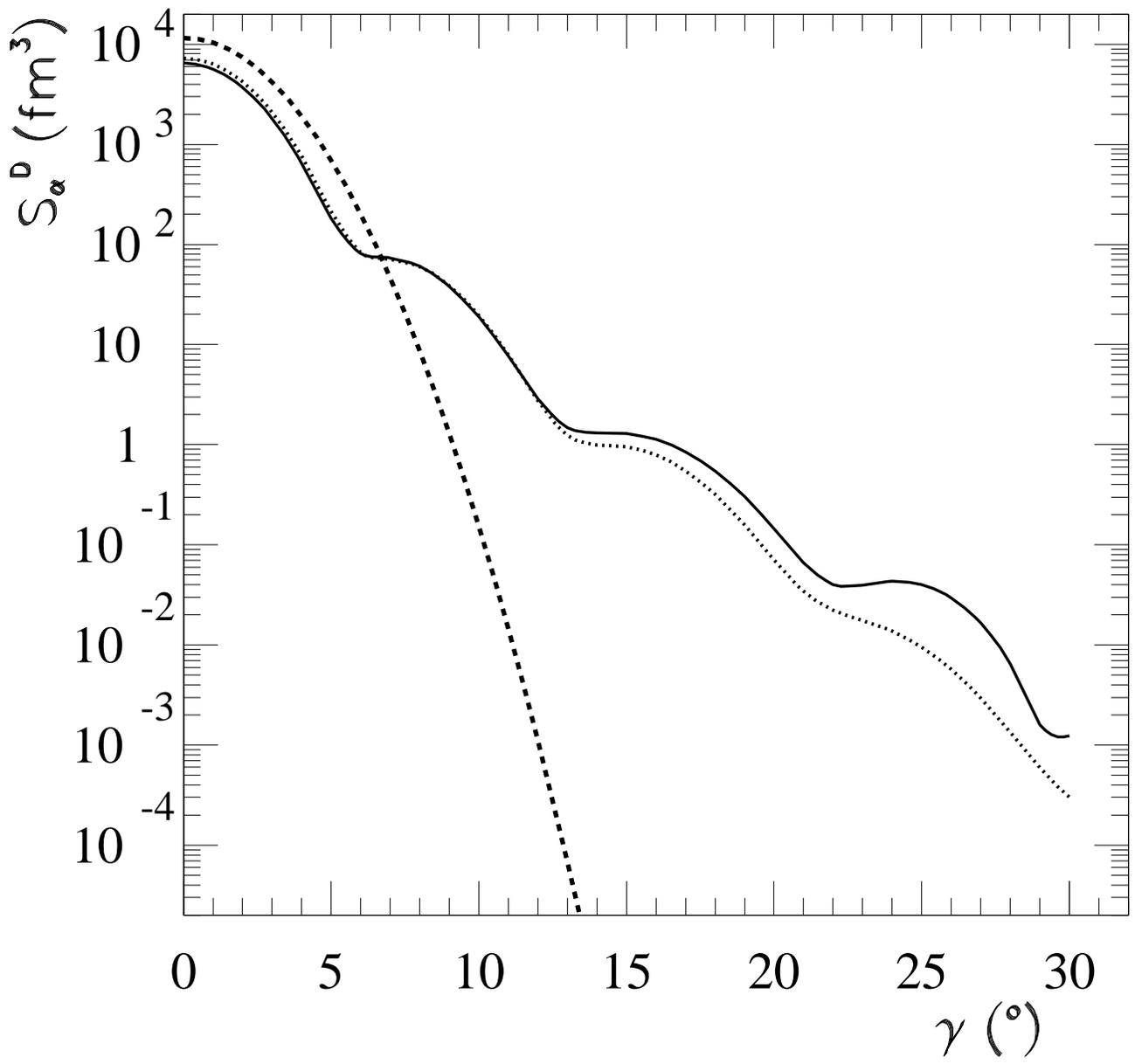
\end{document}